\documentclass[iop]{emulateapj}
\usepackage{natbib, color}
\usepackage[normalem]{ulem}
\usepackage{epstopdf}
\usepackage{graphicx}

\definecolor{red}{rgb}{1.0,0.0,0.0}

\newcommand{\Mj}[1]{$M_\mathrm{Jup}$}

\shorttitle{Architecture of the HD~95086 planetary system}
\shortauthors{Rameau J. et al.}

\begin{document}

\title{Constraints on the architecture of the HD~95086 planetary system with the Gemini Planet Imager}

\author{
Julien Rameau\altaffilmark{1},
Eric L. Nielsen\altaffilmark{2,3},
Robert J. De Rosa\altaffilmark{4},
Sarah C. Blunt\altaffilmark{2,5},
Jenny Patience\altaffilmark{6},
Ren\'{e} Doyon\altaffilmark{1},
James R. Graham\altaffilmark{4},
David Lafreni\`{e}re\altaffilmark{1},
Bruce Macintosh\altaffilmark{3},
Franck Marchis\altaffilmark{2},
Vanessa Bailey\altaffilmark{3},
Jeffrey K. Chilcote\altaffilmark{7},
Gaspard Duchene\altaffilmark{4,8},
Thomas M. Esposito\altaffilmark{4},
Li-Wei Hung\altaffilmark{9},
Quinn M. Konopacky\altaffilmark{9},
J\'{e}r\^{o}me Maire\altaffilmark{7},
Christian Marois\altaffilmark{10,11},
Stanimir Metchev\altaffilmark{12,13},
Marshall D. Perrin\altaffilmark{14},
Laurent Pueyo\altaffilmark{14},
Abhijith Rajan\altaffilmark{6},
Dmitry Savransky\altaffilmark{15},
Jason J. Wang\altaffilmark{4},
Kimberly Ward-Duong\altaffilmark{6},
Schuyler G. Wolff\altaffilmark{16,14},
S. Mark Ammons\altaffilmark{17},
Pascale Hibon\altaffilmark{18},
Patrick Ingraham\altaffilmark{19},
Paul Kalas\altaffilmark{2,4},
Katie M. Morzinski\altaffilmark{20},
Rebecca Oppenheimer\altaffilmark{21},
Fredrik T. Rantakyear\"{o}\altaffilmark{22},
\and Sandrine Thomas\altaffilmark{18}
}

\altaffiltext{1}{Institut de Recherche sur les Exoplan\`{e}tes, D\'{e}partment de Physique, Universit\'{e} de Montr\'{e}al, Montr\'{e}al QC H3C 3J7, Canada}
\altaffiltext{2}{SETI Institute, Carl Sagan Center, 189 Bernardo Avenue, Mountain View, CA 94043, USA}
\altaffiltext{3}{Kavli Institute for Particle Astrophysics and Cosmology, Stanford University, Stanford, CA 94305, USA}
\altaffiltext{4}{Astronomy Department, University of California, Berkeley, CA 94720, USA}
\altaffiltext{5}{Department of Physics, Brown University, Providence, RI 02912, USA}
\altaffiltext{6}{School of Earth and Space Exploration, Arizona State University, PO Box 871404, Tempe, AZ 85287, USA}
\altaffiltext{7}{Dunlap Institute for Astronomy and Astrophysics, University of Toronto, Toronto, ON, M5S 3H4, Canada}
\altaffiltext{8}{Universit\'{e} Grenoble Alpes / CNRS, Institut de Plan\'{e}tologie et d'Astrophysique de Grenoble, 38000 Grenoble, France}
\altaffiltext{9}{Department of Physics and Astronomy, University of California Los Angeles, 430 Portola Plaza, Los Angeles, CA 90095, USA}
\altaffiltext{10}{Department of Physics and Astronomy, University of Victoria, 3800 Finnerty Road, Victoria, BC, V8P 5C2, Canada}
\altaffiltext{11}{National Research Council of Canada Herzberg, 5071 West Saanich Road, Victoria, BC V9E 2E7, Canada}
\altaffiltext{12}{Department of Physics and Astronomy, Centre for Planetary Science and Exploration, The University of Western Ontario, London, ON N6A 3K7, Canada}
\altaffiltext{13}{Department of Physics and Astronomy, Stony Brook University, 100 Nicolls Road, Stony Brook, NY 11790, USA}
\altaffiltext{14}{Space Telescope Science Institute, 3700 San Martin Drive, Baltimore, MD 21218, USA}
\altaffiltext{15}{Sibley School of Mechanical and Aerospace Engineering, Cornell University, Ithaca NY 14853}
\altaffiltext{16}{Physics and Astronomy Department, Johns Hopkins University, Baltimore MD, 21218, USA}
\altaffiltext{17}{Lawrence Livermore National Laboratory, L-210, 7000 East Avenue, Livermore, CA 94550, USA}
\altaffiltext{18}{European Southern Observatory, Alonso de Cordova 3107, Casilla 19001, Santiago, Chile}
\altaffiltext{19}{Large Synoptic Survey Telescope, 950 N. Cherry Ave, Tucson AZ 85719, USA}
\altaffiltext{20}{Steward Observatory, 933 N. Cherry Avenue, University of Arizona, Tucson, AZ 85721, USA}
\altaffiltext{21}{American Museum of Natural History, New York, NY 10024, USA}
\altaffiltext{22}{Gemini Observatory, Casilla 603, La Serena, Chile}

\begin{abstract}
We present astrometric monitoring of the young exoplanet HD~95086~b obtained with the Gemini Planet Imager between 2013 and 2016. A small but significant position angle change is detected at constant separation; the orbital motion is confirmed with literature measurements. Efficient Monte Carlo techniques place preliminary constraints on the orbital parameters of HD~95086~b. With $68~\%$ confidence, a semimajor axis of $61.7^{+20.7}_{-8.4}$~au and an inclination of $153\fdg0^{+9.7}_{-13.5}$ are favored, with eccentricity less than $0.21$. Under the assumption of a co-planar planet-disk system, the periastron of HD~95086~b is beyond $51~$au with $68~\%$ confidence. Therefore HD~95086~b cannot carve the entire gap inferred from the measured infrared excess in the SED of HD~95086. We use our sensitivity to additional planets to discuss specific scenarios presented in the literature to explain the geometry of the debris belts. We suggest that either two planets on moderately eccentric orbits or three to four planets with inhomogeneous masses and orbital properties are possible. The sensitivity to additional planetary companions within the observations presented in this study can be used to help further constrain future dynamical simulations of the planet-disk system.

\end{abstract}

\keywords{planetary system - planet-disk interactions - astrometry - stars: individual (HD 95086) }

\section{Introduction}
Unlike for the majority of extrasolar planets detected through indirect techniques, the orbital periods of planets detected through direct imaging range from decades (e.g., \citealt{Lafreniere:2009,Chauvin:2012}) to thousands of years (e.g., \citealt{Bailey:2014,Naud:2014jx}). So far, only five directly imaged systems have had the orbital parameters of their planets constrained (e.g., \citealp{Beust:2014dj,MillarBlanchaer:2015ha,DeRosa:2015jl,Sallum:2015ej,Zurlo:2016}). In the case of $\beta$~Pictoris, the astrometric measurements cover almost half of a complete orbit, while for the others the astrometric measurements only cover a small fraction of the total orbit. Orbital constraints derived from small astrometric arcs can still be valuable. In the case of Fomalhaut, revealing the high eccentricity of planet b \citep{Kalas:2013hp,Beust:2014dj} led to the need for additional planet(s) to explain the dynamical history of the system \citep{Faramaz:2015}. Dynamical simulations of the HR~8799 system might set upper limits on the masses of the four planets, above which the system would be dynamically unstable at the age of the star \citep[e.g.,][]{Fabrycky:2010, Gozdziewski:2014}.

Orbital configuration of planetary systems can also offer insight into the interactions between planets and circumstellar material. In the case of HR~8799, a three component debris disk is required to model its spectral energy distribution (SED) and resolved images with a dust-free gap \citep{Su:2009ig, Matthews:2013fd}. The semimajor axis of the four planets are both consistent with the location and geometry of this gap, evidence suggesting that the imaged planets are responsible for gravitationally sculpting the disk. This dust distribution is analogous to that seen in the HD~95086 system. HD~95086 ($1.7~M_\odot$, $17\pm4~$Myr, $90.4\pm3.4~$pc) has a large infrared excess \citep{Rizzuto:2011gs,Chen:2012ki} modelled by either two or three component disk \citep{Moor:2013bg,Su:2015ju}, and marginally resolved with \textit{Herschel} \citep{Moor:2013bg}. At a projected separation of 56~AU, intermediate to the two populations of dust, a planetary companion, HD~95086~b, with a model-dependant mass of $4.4\pm0.8$~$M_{\rm Jup}$ \citep{DeRosa:2016}, was discovered by direct imaging \citep{Rameau:2013dr}.

HD~95086~b was imaged during the commissioning of the Gemini Planet Imager (GPI; \citealp{Macintosh:2014js}) in late 2013 \citep{Galicher:2014er}. In this letter, we present astrometric measurements of HD~95086~b obtained with GPI at Gemini South Observatory, Chile, between 2014 and 2016, in which significant orbital motion is measured. By combining our new astrometric measurements and a reanalysis of those presented in \citet{Galicher:2014er}, with previous measurements from \citet{Rameau:2013ds}, we present the first constraints on the orbital parameters of HD~95086~b and discuss implications of the system architecture and properties of any additional planet.

\section{Observations and Data Reduction}

\begin{deluxetable*}{ccccccccc}
\tabletypesize{\footnotesize}
\tablecaption{Observations and Orbital Parameters of HD~95086~\lowercase{b}}
\tablewidth{0pt}
\tablehead{
\colhead{UT Date} &  \colhead{Instrument/} & $T_{\rm int}$ & Field &  Plate Scale & Position Angle & $\rho$ & $\theta$ & \colhead{Ref.}\\
& Filter & (min) & Rot. (deg) & (mas~px$^{-1}$) & Offset (deg) & (mas) & (deg) & }
\startdata
2012 Jan 12 & NaCo/$L\,'$ & $52.0$	& $24.5$ & $27.1\pm0.06$ &		$0.38\pm0.02$ &	$624\pm8$ &		$151.8\pm0.8$ &			   (1)\\
2013 Mar 14	& NaCo/$L\,'$ & $54.0$ & $31.4$ &	$27.1\pm0.04$ &	    $0.45\pm0.09$ &	$626\pm13$ &		$150.7\pm1.3$ &		   (1)\\
2013 Jun 27 & NaCo/$L\,'$ & $151.7$	& $29.3$ & $27.1\pm0.2$ &		$0.48\pm0.1$ &	$600\pm11$ &		$150.9\pm1.2$ &		   (1)\\
2013 Dec 10	& GPI/$K_1$ & $33.4$ & $11.7$ &	$14.166\pm0.007$ &		$-0.1\pm0.13$	& $619\pm5$	&	$150.9\pm0.5$	&   (2)\tablenotemark{{\it a}}\\
2013 Dec 11	& GPI/$H$ & $41.6$ & $15.0$ &	$14.166\pm0.007$ &		$-0.1\pm0.13$	& $618\pm11$	&	$150.3\pm1.1$	&	(2)\tablenotemark{{\it a}}\\
2014 May 13	& GPI/$K_1$ & $34.0$ & $16.9$ &	$14.166\pm0.007$ &		$-0.1\pm0.13$ &	$618\pm8$	&	$150.2\pm0.7$	&   (2)\\
2015 Apr 06	& GPI/$K_1$ & $41.7$ &	$37.7$ & $14.166\pm0.007$ &		$-0.1\pm0.13$ &	$622\pm7$	&	$148.8\pm0.6$	& (2)\\ 
2015 Apr 08	& GPI/$K_1$ & $99.4$ & $39.0$ &	$14.166\pm0.007$ &		$-0.1\pm0.13$	& $622\pm4$	&	$149.0\pm0.4$	&   (2)\\
2016 Feb 29	& GPI/$H$ &	$38.0$ & $16.9$ & $14.166\pm0.007$ &		$-0.1\pm0.13$ &	$621\pm5$	&	$147.8\pm0.5$	& (2)\\
2016 Mar 06	& GPI/$H$ &	$78.0$ & $47.6$ & $14.166\pm0.007$ &		$-0.1\pm0.13$ &	$620\pm5$	&	$147.2\pm0.5$	& (2)\\
\noalign{\smallskip}\hline
\hline
\multicolumn{9}{c}{ }\\
\multicolumn{3}{c}{Parameter} & Unit  & ${\mathcal L}_{\rm max}$ & Median & $68~\%$ CI & $95~\%$ CI & $i$ prior \\
\noalign{\smallskip}\hline\noalign{\smallskip}
\multicolumn{3}{c}{Semimajor axis ($a$)} & au &  50.5 & 61.7 & 53.3--82.4 & 44.8--137 &$\cos(i)$ \\
\multicolumn{3}{c}{}                    &    & 49.3 & 59.7 & 52.7--72.0 & 44.5--105.5 & $155\pm5^\circ$\\
\multicolumn{3}{c}{Eccentricity ($e$)}  & -       & 0.019 & 0.135 & $<$0.205 & $<$0.443 & $\cos(i)$ \\
\multicolumn{3}{c}{}                    &  & 0.168 & 0.124 &  $<$0.187 & $<$0.397 & $155\pm5^\circ$\\
\multicolumn{3}{c}{Inclination ($i$)}   & deg  & 146.0 & 153.0 & 139.5--162.7 & 126.1--174.7 & $\cos(i)$\\
\multicolumn{3}{c}{}                    &  &  149.2 & 155.7 & 150.9-160.6 & 146.2--165.5 & $155\pm5^\circ$ \\
\multicolumn{3}{c}{Argument of periastron ($\omega$)} & deg & 39.7 & 92.5 & 27.3--153.7 & 3.73--176.2 & $\cos(i)$\\
\multicolumn{3}{c}{}                    &  & 99.2 & 92.6 & 28.1--153.0 & 3.92-176.1 & $155\pm5^\circ$ \\
\multicolumn{3}{c}{Position angle of nodes ($\Omega$)} & deg &  329.7 & 83.8 & 38.2--146.7 & 5.48--174.7 & $\cos(i)$\\
\multicolumn{3}{c}{}                    &  &  2.5 & 89.8 & 32.4--149.4 & 4.76-175.3 & $155\pm5^\circ$\\
\multicolumn{3}{c}{Epoch of periastron ($T_0$)} & - &  2172.37 & 2155.16 & 2040.26-2367.95 & 2005.35-2809.37 & $\cos(i)$ \\
\multicolumn{3}{c}{}                     &  &  2202.44 & 2151.83 & 2038.27--2337.07 & 2005.12-2580.66 & $155\pm5^\circ$ \\
\multicolumn{3}{c}{Period ($P$)}        & year &  257.1 & 370.9 & 296.8--572.6 & 227.9-1230.9 & $\cos(i)$\\
\multicolumn{3}{c}{}                    &  &  260.0 & 352.8 & 291.2--469.2 & 225.5--834.1 & $155\pm5^\circ$
\enddata
\tablenotetext{a}{These data were published in \citet{Galicher:2014er} but re-analyzed in this work.}
\tablenotetext{}{\textbf{References.} (1) \citet{Rameau:2013ds}; (2) This work }
\label{tab:obs}
\end{deluxetable*}

\begin{figure}[th]
  \centering
  \includegraphics[width=0.5\textwidth]{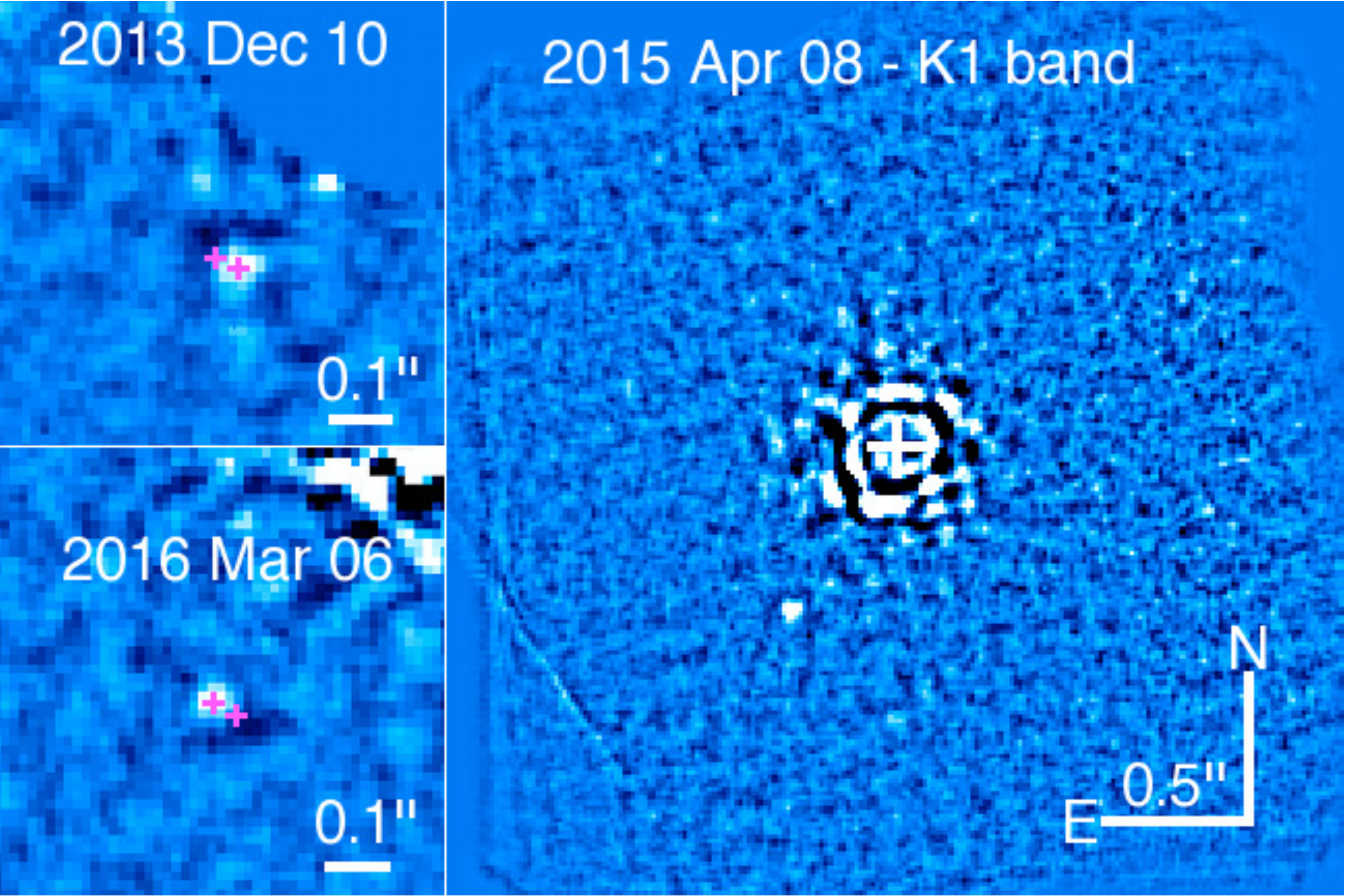}
    \caption{(Left:) Zoom-in images on HD~95086~b obtained with GPI at the first and last epochs. The magenta crosses show the measured positions (for clarity, the size of the symbol is not representative of the precision). Significant orbital motion is detected within the GPI data. (Right:) Deepest image obtained on HD~95086 with GPI at $K_1$ on 2015 April 08.}
\label{fig:image}
\end{figure}

Observations of HD~95086~b were made with GPI at four different epochs between December 2013 and March 2016 with either the $H$ (1.5--1.8~\micron) or $K_1$ (1.9--2.2~\micron) filters (program IDs GS-ENG-GPI-COM, GS-2015A-Q-501, GS-2015B-Q-500, and GS-2016A-Q-12). HD~95086 was also observed on 2014 March 24 and 26 (program ID GS-ENG-GPI-COM) but the image quality was not sufficient to detect the planet. The observing sequences consisted of an observation of an argon arc-lamp to calibrate the offset between the position of the microspectra at the target elevation and at zenith caused by instrument flexure \citep{Wolff:2014}, followed by between 30 and 100 minutes of on-source integration, centered at meridian passage to maximize the field rotation for angular differential imaging \citep{Marois:2006df}. A log of the observations is given in Table \ref{tab:obs}.

The GPI observations on 2013 December were first published by \citet{Galicher:2014er} but were reprocessed in this work for a uniform set of measurements, owing to an improvement of the spot registration and astrometric calibration.The raw GPI images were reduced with the Data Reduction Pipeline (DRP, v1.3.0\footnote{\tt http://docs.planetimager.org/pipeline/}) to produce $(x, y, \lambda)$ calibrated datacubes. The DRP was also used to measure the satellite spot positions for precise image registration \citep{Wang:2014}. The reduced images were filtered with an unsharp mask ($15\times15~$pixels), the speckles were then subtracted using LOCI \citep{Lafreniere:2007}, with $dr=5$~pixels, $N_A=300~$times the full-width-at-half-maximum (FWHM), $g=1$, and $N_{\delta}=0.75$ or $1~$FWHM depending on the amount of field rotation. The residual images were finally rotated so that north was aligned with the vertical axis, combined with a trimmed mean ($10~\%$), and collapsed to build a broadband final image. 

The astrometry of HD~95086~b was extracted at all epochs following \citet{Marois:2010b, Lagrange:2010}. A model point spread function (PSF) was built from the spots and injected into the data at an initial position. The amoeba-simplex optimization method \citep{Nelder:1965in} was used to iterate over the position and flux to minimize the residuals in a wedge of $\Delta r\times\Delta \theta=2\times2~$FWHM. Errors were calculated by injecting fake planets at the same separations but twenty positions angles uniformly distributed between $90^\circ$ and $270^\circ$ away from planet b and repeating the same exercise. Final uncertainties on the relative companion position were computed by adding quadratically the errors from the measurements, the spot registration, and the plate scale and position angle offset (orientation of North with respect to the vertical axis). For the GPI measurements we used the same astrometric solution as described in \citet{DeRosa:2015jl} to convert the image position $(x,y)$ into an on-sky separation and position angle. The results of this analysis are presented in Table \ref{tab:obs}. Measurements from the literature were also compiled in this study \citep{Rameau:2013dr, Rameau:2013ds}. They were obtained with VLT/NaCo at $L\,'$ (3.5--4.1~\micron) at three different epochs between January 2012 and June 2013.

\section{Constraints on the orbit of HD 95086 b}

\begin{figure*}[ht]
\centering
\includegraphics[width=\textwidth]{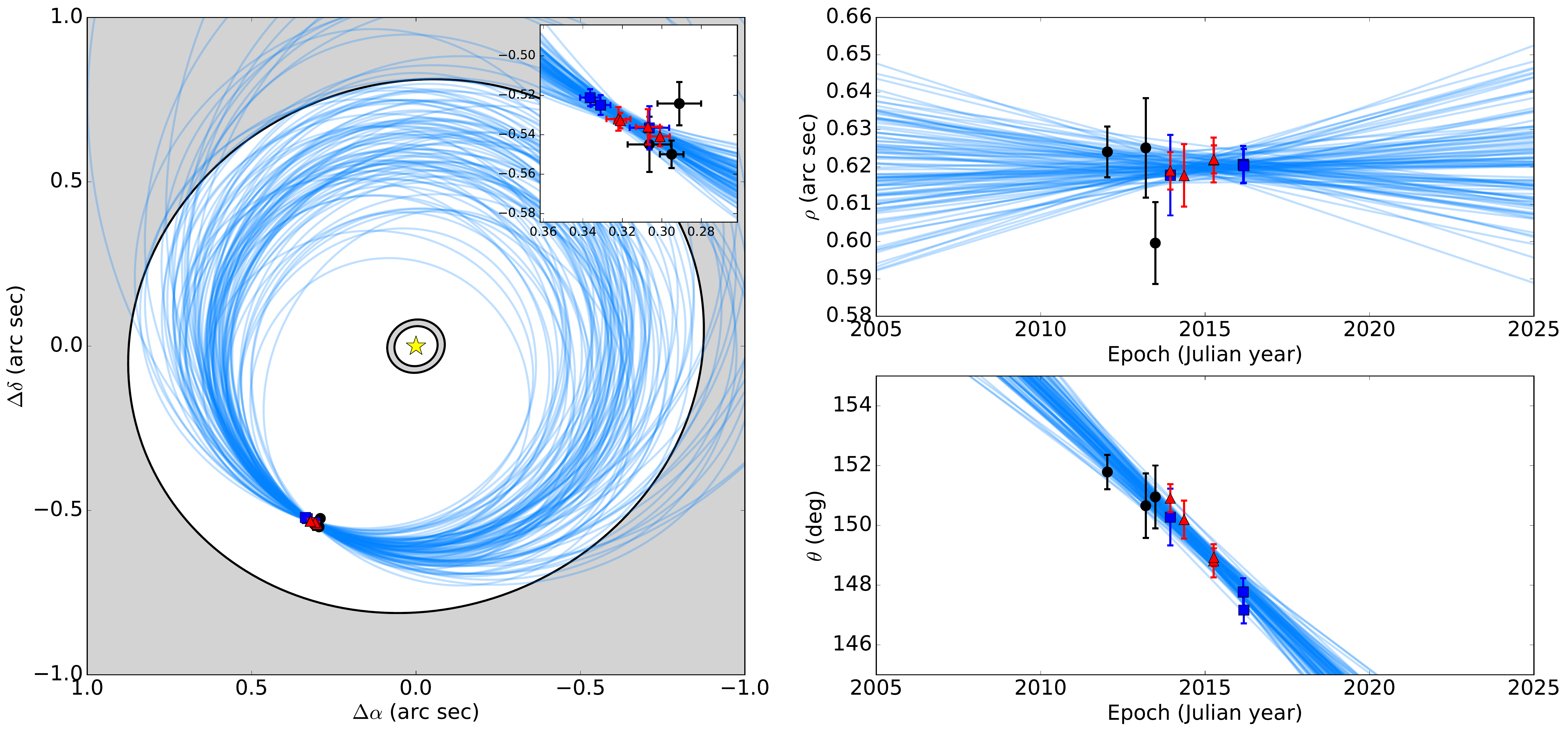}
\caption{(Left:) Schematic diagram of the HD~95086 system in the sky plane. The astrometric measurements of HD~95086~b are plotted (black circles - NaCo $L\,'$, red triangles - GPI $K_1$, blue squares - GPI $H$), as well as a hundred representative orbital fits randomly drawn from the rejection sampling analysis using the inclination-restricted ($i=155\pm5^\circ$) prior. The inner and outer dust rings are indicated as the gray shaded regions, based on the average values from \citet{Su:2015ju}. For both dust rings, an inclination of $i=155^{\circ}$ and a position angle of $110^{\circ}$ were assumed \citep{Su:2015ju}. For clarity, the astrometric measurements are also shown within an inset. (Right:) The separation (top right) and position angle (bottom right) of HD~95086~b measured between 2013 and 2016. Symbols and lines are as in the left panel.}
\label{fig:orbit_plot}
\end{figure*}

\begin{figure*}[ht]
\centering
\includegraphics[width=\textwidth]{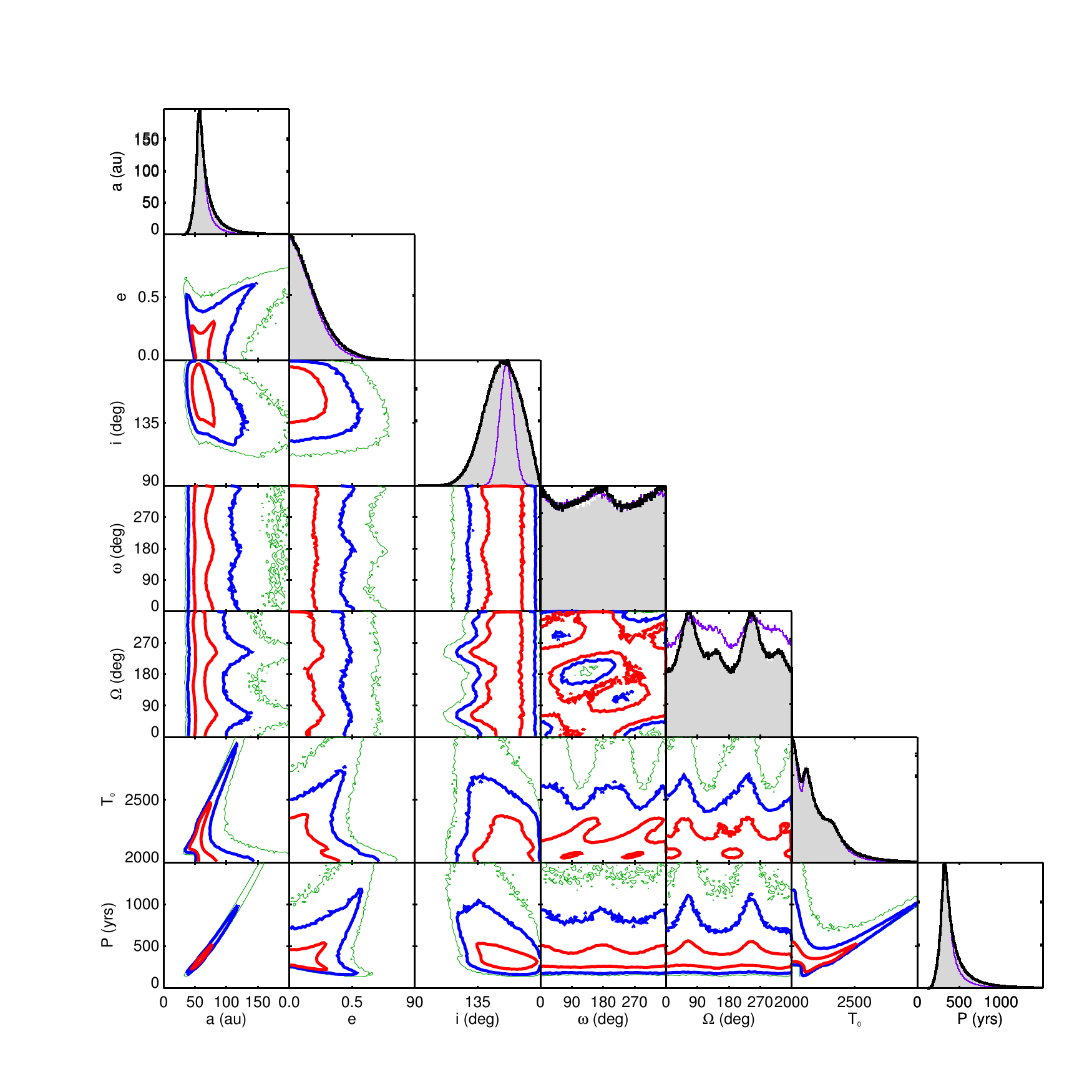}
\caption{Posterior distributions (black histograms) of the orbital parameters of HD~95086~b as fit by our rejection sampling technique, assuming a uniform prior in $\cos i$. Off-diagonal elements give the covariance between orbital elements, with red, blue, and green contours showing where $68~\%$, $95~\%$, and $99.7~\%$ of the orbits are contained, respectively.   Also plotted in the on-diagonal elements are an MCMC fit to the same data with the same priors (gray shaded histogram) and the results from a similar fit using the rejection sampling technique but with the inclination-restricted ($i=155\pm5^\circ$) prior (thin purple histogram).}
\label{fig:triangle}
\end{figure*}

The astrometry of HD~95086~b given in Table~\ref{tab:obs} showed a roughly constant separation and a decreasing position angle with time, consistent with face-on, circular orbital motion (see Figures \ref{fig:image} and \ref{fig:orbit_plot}). The orbital motion was still clear within the GPI data sets which covered more than two years. As in \citet{DeRosa:2015jl} we fitted the orbit using a rejection sampling technique that is more computationally efficient than traditional Markov chain Monte Carlo (MCMC) techniques at generating posterior parameter distribution for short orbital arcs.  Briefly, this technique sampled the posterior probability density function of orbital parameters by generating orbits from prior distributions, scaling the semimajor axis and rotating the position angle of nodes to fit one of the observational epochs, rotating through each epoch.  Final orbits were rejected or accepted by comparing the probability ($P \propto {\rm e}^{- \chi^2/2}$) of the remaining astrometric epochs to a uniform random variable.  This method, introduced in \citet{Blunt:2016}, will be described more in depth in Blunt et al. (2016, in preparation).  To fit HD~95086~b we used priors that are uniform in $\log a$, $\cos i$, $\omega$, $\Omega$, and $T_0$, and a linearly decreasing prior in $e$ fit to radial velocity planets \citep{Nielsen:2008}. Based on the inclination of the outer disk of $i=25\pm5^\circ$ (without knowledge of the rotation direction both $25^\circ$ and $180-25=155^\circ$ are possible) \citep{Moor:2013bg,Su:2015ju}, we also ran a fit to the astrometry with a restricted prior on the inclination. A pair of Gaussians centered on $i=25^{\circ}$ (counter-clockwise motion) and $180-25=155^{\circ}$ (clockwise motion), both with $\sigma=5^{\circ}$, was used as the prior on the inclination.

Figure~\ref{fig:triangle} plots the posterior distributions of the orbital parameters of HD~95086~b.  Overall the posteriors distributions correspond to face-on, circular orbits. Table~\ref{tab:obs} gives, for the two inclination priors, the most likely (${\mathcal L}_{\rm max}$), median, and 68\% and 95\% confidence intervals (CI) for each orbital parameter. For the inclination-restricted fit, the posterior distributions are not significantly changed, although the median of the inclination distribution was shifted to slightly higher values due to the shape of the prior. Our finding of a low eccentricity was relatively independent of our choice of the prior on either the inclination and eccentricity, currently $e<0.44$ at $95~\%$ confidence using the linear prior in $e$, but switching to a uniform prior still resulted in $e<0.56$ at $95~\%$ confidence. Fitting the data points without the outlier from 2013 June 27 did not affect our conclusions about the semimajor axis, eccentricity, and inclination of the orbits.

We confirmed the reliability of the rejection sampling fits with the same Metropolis-Hastings MCMC orbit fitting used in \citet{Nielsen:2014}, using the same priors on parameters as above.  As in \citet{DeRosa:2015jl} we found excellent agreement between the two methods, as shown in the diagonal elements of Figure~\ref{fig:triangle}.

In addition to these Monte Carlo techniques, the method for constraining orbital parameters over short orbital arcs presented in \citet{Pearce:2015je} was applied to the astrometry of HD~95086~b. The angle between the projected separation and velocity vectors was calculated as $\varphi=96.7^{+9.4}_{-9.2}~$deg, and a value of the dimensionless parameter $B$ of $0.52^{+0.20}_{-0.16}$. Comparing these to the minimum inclination and eccentricity contours of \citet{Pearce:2015je}, the orbital parameters of HD~95086~b can only be constrained to $e \ge 0.15_{-0.15}^{+0.31}$, and $i\le59.1^{+9.6}_{-13.8}$~deg (corresponding to $i\ge120.9^{+13.8}_{-9.6}$~deg). In each case, uncertainties on the measured positional offset of HD~95086~b were propagated in a Monte Carlo fashion. While these limits are consistent with the values in Table~\ref{tab:obs}, they are significantly less constraining.

\section{Constraints on additional planets}
\begin{figure*}
\centering
\includegraphics[width=0.5\textwidth]{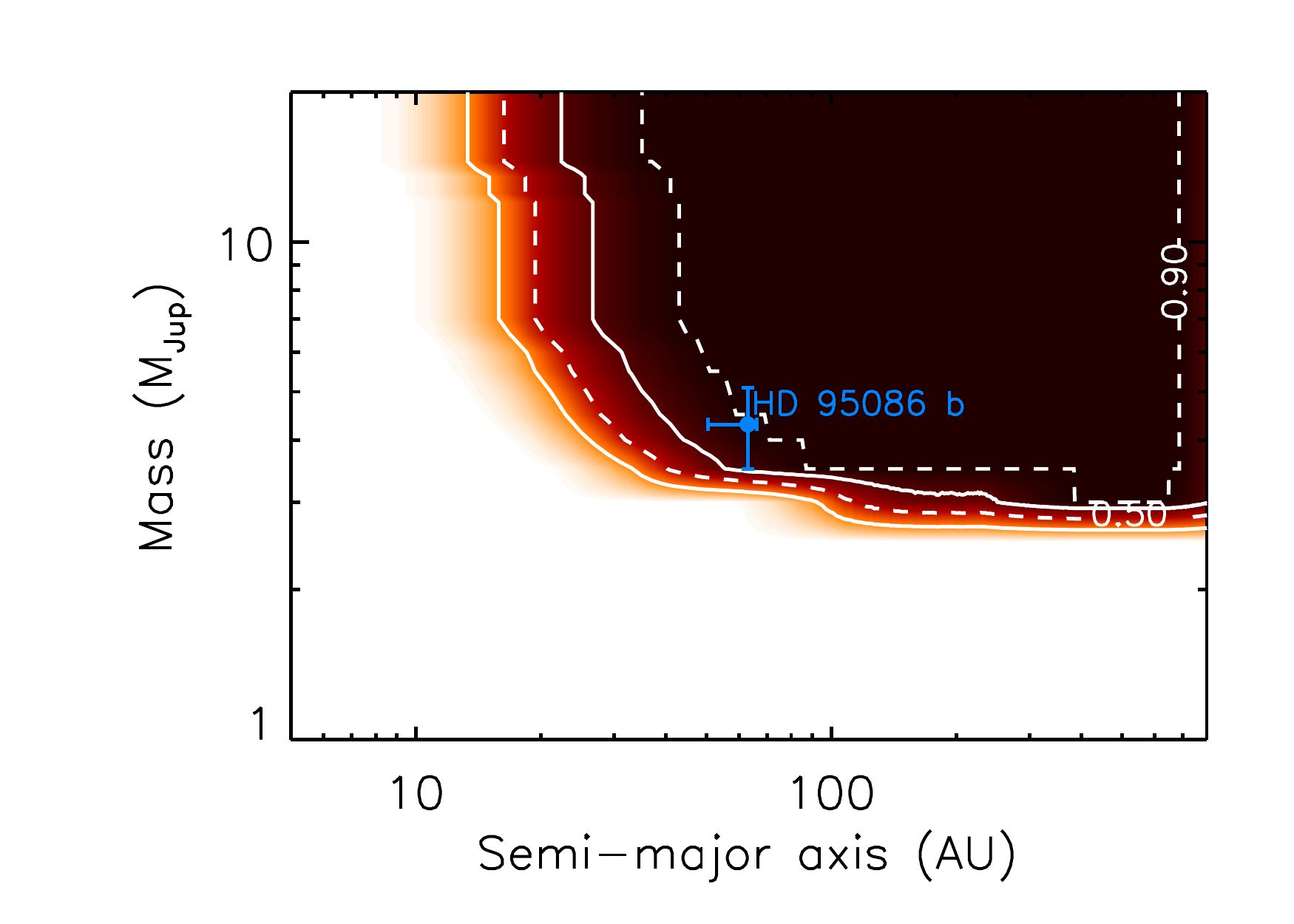}\includegraphics[width=0.5\textwidth]{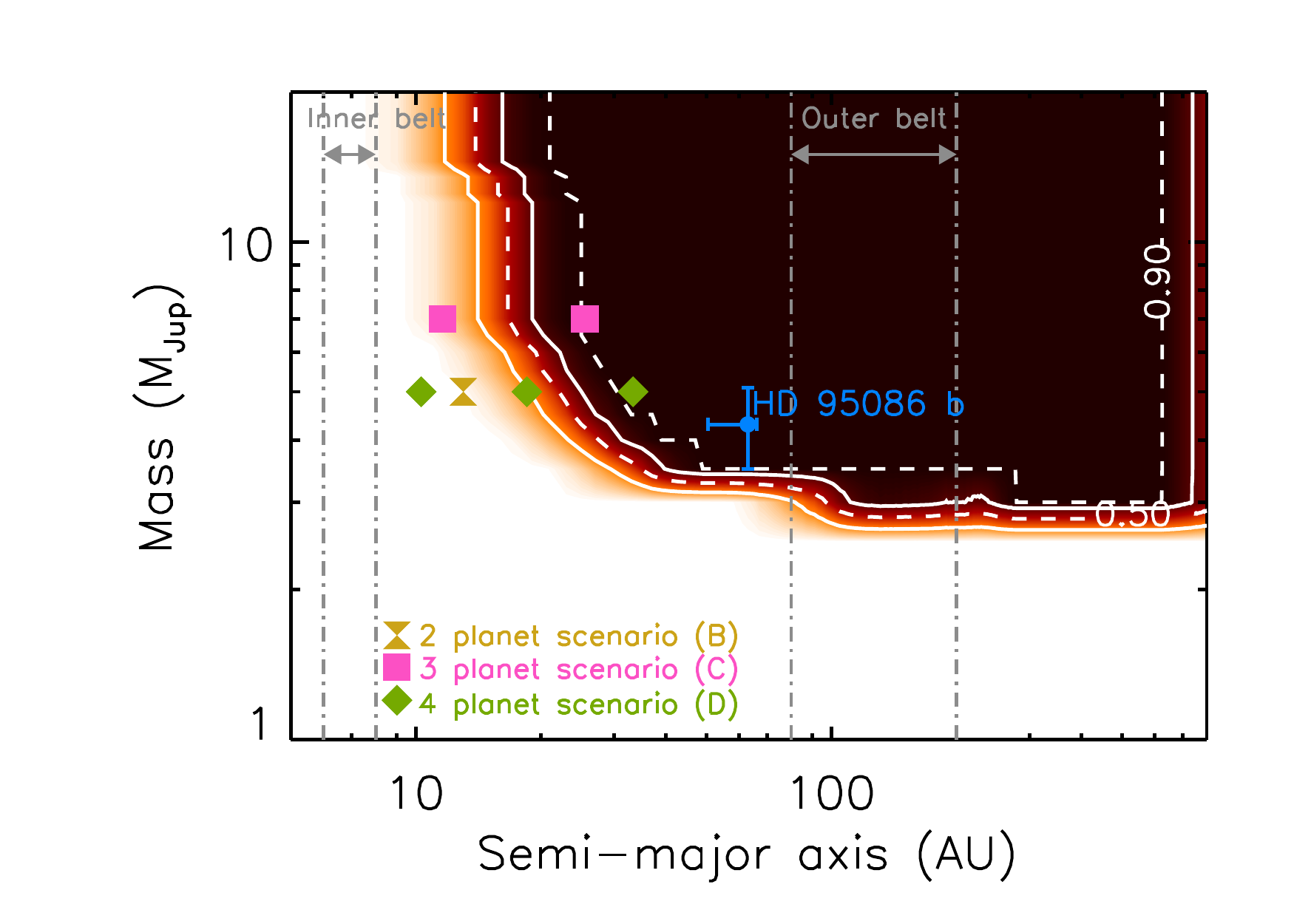}
\caption{Completeness maps built from the  $5~\sigma$ detection limits of NaCo at $L\,'$ taken in 2012 and of GPI at $K_1$ in 2015. Luminosity to mass conversion was done from the AMES-COND model for an age of $17~$Myr  \citep{Baraffe:2003bj}. Contours specify the $25, 50, 70,$ and $90\%$ detection probabilities. (Left:) Case of the prior uniform in $\cos(i)$. (Right:) Case of the inclination-restricted prior ($i=155\pm5^\circ$) with $e<0.7$. The average location of the inner and outer disks from \citet{Su:2015ju} are indicated with the vertical dashed-dotted lines. The planet properties in the scenarios of \citet{Su:2015ju} are shown with different symbols. The planets in the scenarios C and D orbit with $e<0.1$ and $e=0$ respectively. Therefore, they would have been more easily detected (by 10--20~\%) compared to that is shown in this plot, which includes higher eccentricities. In both panels, HD~95086~b is plotted with a blue circle, based on the $68\%$ confidence interval on the semimajor axis from the inclination-restricted orbital fit given in Table~\ref{tab:obs}.}
\label{fig:limit}
\end{figure*}

The point source sensitivity of the GPI data was estimated by measuring the noise in concentric annuli in the residual LOCI images. The throughput was computed by injecting fake planets in the raw data that were reduced with the same coefficients as the science images. The most sensitive GPI observations were obtained on 2015 April 8 at $K_1$ band (see Figure \ref{fig:image}), and the deepest NaCo $L\,'$ detection limit was taken from \citet{Rameau:2013ds}. The planet-to-star contrast was converted into predicted mass with the AMES-COND \citep{Baraffe:2003bj} model. An optimized version of the Monte Carlo based MESS tool \citep{Bonavita:2012} was used to generated random on-sky positions of planets in order to compute detection probabilities over a separation range of 1--1000~au, with a $2~$au step size, and a planet mass range of 0.5--20~M$_\mathrm{Jup}$, with a $0.5~M_\mathrm{Jup}$ step size. The distributions of the orbital parameters were the same as for the orbit fitting. For each point in the mass--semimajor axis grid, ten thousand orbits were randomly generated, and the fraction of planets which would have been detected in either the GPI or NaCo observations was used as the completeness at that point. The final completeness map built is shown in Figure \ref{fig:limit} (left).

The architecture of the HD~95086 system can now be constrained based on the first estimates of the orbital parameters of planet b and on the detection limits reached with the current ensemble of observations. \citet{Su:2015ju} suggested that the belts that produce the dust properties inferred from the SED and the \textit{Herschel} images are separated by a dust-free gap from $\sim$8 to $\sim$80~au. They proposed four non-exhaustive scenarios to explain the large size of the gap. While there are currently large uncertainties in the exact location of the dust rings, and a large number of alternative multiple planet configurations which may explain the disk gap, the current observational constraints can be used to explore the four specific scenarios presented in \citet{Su:2015ju}. By assuming a system in which the disk and planet(s) are co-planar, the orbital parameters of HD~95086~b, and the associated completeness for additional companion (Figure \ref{fig:limit}, right panel), allows us to:
\begin{itemize}
    \item rule out scenario A in which the planet b would be responsible for clearing the entire gap with an eccentricity of $\sim$0.7. Based on the orbit fit presented in Table \ref{tab:obs}, an eccentricity of $e>0.40$ for planet b can be excluded at the 95\% confidence level, and therefore planet b is unlikely to alone account for the gap. Moreover, \citet{Su:2015ju} argued that the outer disk might have $e<0.25$. This morphology is consistent with the probable orbit of the planet.
    \item neither validate nor eliminate scenario B in which planet b resides on a moderately eccentric orbit ($\sim$0.3), within the allowed range of orbital parameters presented in Table \ref{tab:obs}, and with a semimajor axis of $\sim$40~au, at the edge of the $95\%$ confidence interval. Another inner planet with moderate eccentricity at $\sim$16~au is required but is out of reach of our observations. Other configurations proposed by \citet{Su:2015ju} with a more massive inner planet on a higher eccentric orbit and planet b with reduced eccentricity are unconstrained by our observations.
    \item reconsider scenario C in which two additional $7~\mathrm{M}_\mathrm{Jup}$ planets are needed at $\sim$12 and $\sim$26~au with low eccentricities. An eccentricity of $\sim$0.1 and a semimajor axis of $\sim$56~au for planet b as suggested by this model are permitted by the orbital fitting. However, the second planet in this scenario at $26$~au with a mass of $7~M_\mathrm{Jup}$ and $e<0.3$ would have been detected in our observations. A conservative $75\%$ threshold was reached at this separation at $4~M_\mathrm{Jup}$, or at $25~$au with the expected mass, thus very close to the predicted properties of the second planet. The inner planet remained undetectable. 
    \item rule out scenario D in which three equal mass ($5~\mathrm{M}_\mathrm{Jup}$) planets on circular orbits at $\sim$11, $\sim$19, and $\sim$34~au populate the gap in addition to planet b. The third $5~M_\mathrm{Jup}$ planet would have been detected at $100\%$ confidence at the predicted position with a null eccentricity.
\end{itemize}

Therefore, in the context of a $\sim$8--80~au gap with coplanar planets with masses derived from the AMES-COND models, the scenario with two planets (B) cannot be ruled out, whereas the scenario with three $7~M_\mathrm{Jup}$ planets (C) is inconsistent with our observations, instead the second planet would require a higher eccentricity with either a lower mass, or smaller semimajor axis in order to be consistent with our observational constraints. The tempting four equal-mass ($5~M_\mathrm{Jup}$) planet scenario (D), analogous to the HR\,8799 system, was rejected based on the non-detection of the third planet. All the scenario might however still be accepted if the predicted masses are wrong by a factor of a few, assuming other evolutionary models \citep[e.g.][]{Spiegel:2012, Marleau:2014}.

As noted previously, the architecture of the debris belts was inferred from the SED and marginally resolved image data of the system and provided dynamical constraints on the location of planets. However, the geometry of a debris disk based solely on SED modelling can be wrong by a factor of a few \citep{Booth:2013,Morales:2013,Pawellek:2014}. Grain properties and dust position are degenerate. Resolved images of the different disk components are necessary to break this degeneracy \citep[e.g.,][]{Lebreton:2012}. This behavior was mentioned in \citet{Su:2015ju} in which the warm and cold belts might be located between $7$ and $8~$au, and between $60$ and $125~$au respectively. The outer edge of the inner belt might also be poorly determined if the emission is dominated by its innermost region, artificially making a more extended inner disk into a narrow and closer-in ring. These uncertainties affect the predicted semimajor axes, eccentricities, and masses of the planets in the aforementioned scenarios. Even if a conservative $75\%$ detection threshold was applied to assess the different hypotheses, a change by a factor of a few in mass or semimajor axis might result in a drastic change in detection probability at the predicted properties of the planets. 

\section{Concluding remarks}
We have presented four epochs of astrometric measurements of the young giant planet HD~95086~b with the Gemini Planet Imager. We were able to detect significant change in the position angle of the planet between the end of 2013 and early 2016 whereas the separation remained similar. By using Monte Carlo methods to sample probable orbits, we placed the first constraints on the orbital parameters of HD~95086~b. The semimajor axis distribution peaks near its median value of $61.7~$au while the eccentricity and inclination were limited to $e<0.21$ and $i=153\fdg0^{+9.7}_{-13.5}~$ respectively at $68~\%$ confidence.

 We also discussed the system architecture in the context of a coplanar planet(s)-debris disk configuration. Based on the preliminary constraints on the orbit of planet b, and on the detection probabilities, we tested the scenarios from \citet{Su:2015ju} to explain the sustainability of the $\sim$8--80~au dust-free gap inferred from SED fitting and resolved sub-mm imaging. Of the four scenarios, only the two planet scenario (B) cannot be excluded based on the observational constraints presented here. The upper limit on the eccentricity of planet b confidently ruled out a single planet scenario (A), and the GPI sensitivity excluded the three and four equal-mass planet scenarios (C and D) by a non-detection of the second and third planet on low eccentricity orbits at $26$ and $34~$au, respectively. The four scenarios presented in \citet{Su:2015ju} are not exhaustive, and  modifications to masses and orbits of additional companions are still possible to sustain the gap, while also being dynamically stable for $17~$Myr.

Due to the degeneracy between the properties of additional planets, extensive exploration of the parameter space is necessary to further assess the architecture of the system beyond the analytical statements made above. Such an analysis will also require the precise geometry of the debris. Resolved images of the outer disk with ALMA, or by polarimetry with GPI or SPHERE \citep{Beuzit:2008}, combined to the detection of the inner disk by interferometry will not only constrain the locations of the inner and outer edges of the gap, but also the eccentricities of the warm and cold belts.

\acknowledgments

We thank Kate Y. L. Su for interesting discussions about the debris system and Virginie Faramaz for her fruitful advices on handling chaotic zones. Based on observations obtained at the Gemini Observatory, which is operated by the Association of Universities for Research in Astronomy, Inc., under a cooperative agreement with the National Science Foundation (NSF) on behalf of the Gemini partnership: the NSF (United States), the National Research Council (Canada), CONICYT (Chile), the Australian Research Council (Australia), Minist\'{e}rio da Ci\^{e}ncia, Tecnologia e Inova\c{c}\~{a}o (Brazil) and Ministerio de Ciencia, Tecnolog\'{i}a e Innovaci\'{o}n Productiva (Argentina). J.R., R.D. and D.L. acknowledge support from the Fonds de Recherche du Qu\'{e}bec. Supported by NSF grants AST-1518332 (R.J.D.R., J.R.G., J.J.W., T.M.E., P.K.), AST-1411868 (B.M., A.R., K.W.D.), and AST-141378 (G.D.). Supported by NASA grants NNX15AD95G/NEXSS and NNX15AC89G (R.J.D.R., J.R.G., P.K., J.J.W., T.M.E.), and NNX14AJ80G (E.L.N., S.C.B., B.M., F.M., M.P.). Portions of this work were performed under the auspices of the U.S. Department of Energy by Lawrence Livermore National Laboratory under Contract DE-AC52-07NA27344 (S.M.A.).

{\it Facility:} \facility{Gemini:South (GPI)}.

\bibliography{refs.bib}


\end{document}